# Vortex, Defect, and Monopole Gauge Fields for Nambu-Goldstone Systems and their Phase Transitions

or

# Theory of Fluctuating Multivalued Fields

H. Kleinert
Institut für Theoretische Physik
Freie Universität Berlin
Arnimallee 14    D - 14195 Berlin

**Abstract.** We survey the various uses of singular gauge fields which render an ideal description of ensembles of line- and surface-like excitations and explain phase transitions with the help of simple quadratic actions. Near a transition, the singular gauge fields can be transformed into ordinary complex fields called disorder fields which differ from Landau's order fields by developing a nonzero expectation value in the high-temperature phase.

## 1. Introduction

Many physical systems possess massless excitations. These dominate their statistical properties at low temperatures. The specific heat, for example has the characteristic temperature behavior

$$C \sim T^D \tag{1}$$

in $D$ space dimensions (see Fig. 1 for solids and Fig. 2 for superfluid $^4$He).

The origin of massless excitations can often be explained by a spontaneously continuous symmetry of the action. Then the excitations are called Nambu-Goldstone (NG) modes. Examples are sound waves in superfluids and solids and magnons in ferromagnets.

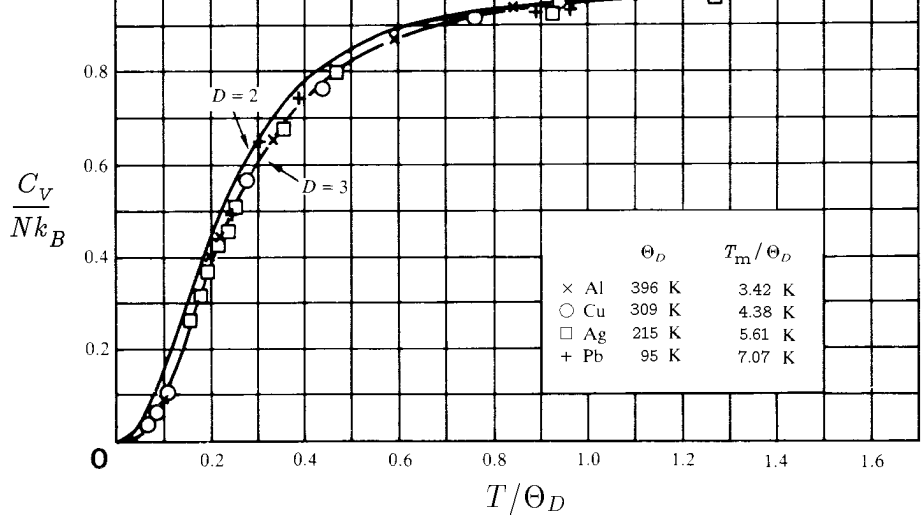

**Figure 1.** The specific heat of various solids. By plotting the data against the ratio $T/\Theta_D$, where $\Theta_D$ is the Debye temperature, the data fall on a universal temperature $T_\mathrm{m}$.

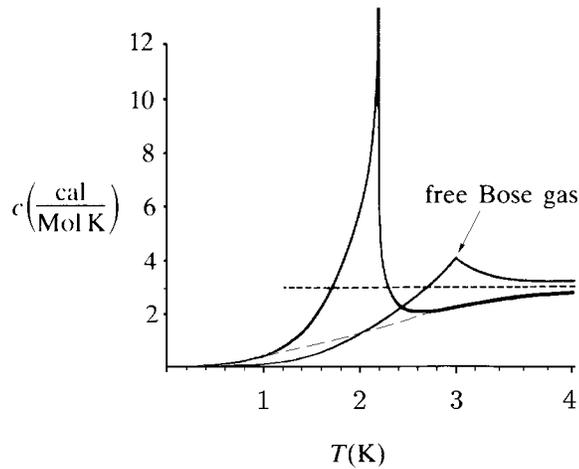

**Figure 2.** The specific heat of superfluid $^4$He. The peak is caused by the superfluid-normal transition. A $T^3$-behavior is clearly visible under the peak.

In superfluid $^4$He, the spontaneously broken symmetry is a $U(1)$ phase symmetry, in which the complex order field $\psi(\mathbf{x})$ goes over into

$$\psi(\mathbf{x}) \to e^{i\alpha}\psi(\mathbf{x}). \tag{2}$$

In solids, the translational symmetry of the atomic position

$$x_i \to x_i + u_i(\mathbf{x}), \tag{3}$$

is spontaneously broken. In ferromagnets, it is the rotational symmetry of the direction $\mathbf{m}$ of the elementary magnets

$$\mathbf{m}(\mathbf{x}) \to R\,\mathbf{m}(\mathbf{x}). \tag{4}$$

which is spontaneously broken.

As the temperature is raised, the $T^D$-behavior goes over into the typical Debye type of behavior. In crystals, it follows a universal function of $T/\Theta_D$, where $\Theta_D$ is the following the Dulong-Petit law which depends only on the number of harmonic degrees of freedom.

This picture is modified by anharmonic effects which become visible at increasing temperatures and which may be neglected in a qualitative discussion. In addition, there are usually massive excitations which, however, are strongly suppressed at low temperatures by a Boltzmann factor $y = e^{-E_a/T}$ called the *fugacity* of the excitations. Here $E_a$ is the *activation energy* defined as the energy of an excitation of infinite wave length, which in relativistic systems is called the *mass* of an excitation.

At a certain temperature $T_c$, which in solids lies above the Debye temperature, in quantum solids and superfluid $^4$He near or below the Debye temperature, the statistical behavior undergoes a dramatic change. The physical properties of the system change in a phase transition. This often happens at energies where the fugacities of massive excitations are very small so that up to the transition, only $NG$-modes should be substantially excited. The dramatic change is caused by the sudden appearance of line-like excitations, whose large activation energies are overcome by a high configurational entropy. In solids, these excitations are defect lines, in superfluids, they are vortex lines. This entropy mechanism mechanism was noted in the fifties by Shockley and Feynman. A line-like excitation of length $l$ with an excitation energy $\epsilon$ per unit length is suppressed by a Boltzmann factor $e^{-\epsilon l/T}$. If the line can bend easily on a length scale $a$, it has roughly $(2D)^{l/\xi}$ possible configurations We can therefore estimate the partition function of an ensemble of such lines with various length as

$$Z \approx \oint dl \, (2D)^{l/\xi} e^{-\epsilon l/T}. \tag{5}$$

The integral converges only

$$T < T_c = \epsilon \xi / \log(2D). \tag{6}$$

Above this temperature, the integral diverges and the ensemble undergoes a phase transition in which the lines proliferate and become infinitely long. We shall call this a *condensation process* of lines.

It is an interesting task to construct a theory which explains the phase transitions in NG-sytems with the help of these line-like excitations without invoking explicitly the non-linear microscopic properties which are the origin of these excitations. The purpose of these lectures is to show that this is possible: There exists a minimal extension of the Debye theory of the harmonic NG-excitations which exhibit a phase transition caused by the condensation of idealized infinitely thin line-like excitations. The important theoretical tool to describe these is a new type of fluctuating fields which I call generically *defect gauge fields*, in the superfluid *vortex gauge field*. In contrast to ordinary fluctuating fields, vortex or defect gauge fields have the useful property that they are capable if explaining the above phase transitions with a quadratic field action. This makes it easy to transform their partition functions to various other forms by quadratic manipulation of functional integrals. By a particular variable change, the partition function can be converted into an interacting field theory of the $|\psi|^4$-type which is called *disorder field theory*.

The main difference with respect to the familiar order field theories of the Landau type is an opposite temperature behavior. In contrast to order fields, disorder fields acquire a nonzero expectation value at high temperatures.

to long-range interactions between the line elements. The interaction can be explained by a local field interaction if the description of the massless NG-modes is changed into a gauge field description. This is always possible. The new gauge fields are called generically *stress gauge fields*. The associated gauge invariance is dual to that of the defect gauge fields.

In this way we arrive at a stress- and defect-gauge invariant theory of Goldstone modes and defect lines, or at an equivalent stress-gauge-invariant theory of disorder fields.

Since NG-modes occur as a result of a spontaneous breakdown of some continuous symmetry, their complex order field description has the form

$$\psi(\mathbf{x}) = e^{-i\theta_a(\mathbf{X})G^a}\psi_0 \tag{7}$$

where $\psi_0$ is a fixed order parameter and $G^a$ is a subset of generators of the associated symmetry group $g$. In the limit of infinite wave lengths, $\theta_a(\mathbf{x})$ becomes a constant and the energy is independent of $\theta_a$— this being the reason for the masslessness the NG-modes. In any order field theory of the Landau type, the so-called London limit selects from all excitations precisely the NG-modes and approximates their field energy harmonically, while freezing all other excitations. The gradient energy of the order field gives rise to a gradient energy of the NG-modes in the London limit

$$\mathcal{A}_{\mathrm{NG}} \propto \int d^D x |\partial e^{-i\theta_a(\mathbf{x})G^a}|^2. \tag{8}$$

It is a nontrivial to translate this energy explicitly the $\theta_a(\mathbf{x})$-fields. There always exist points in the field space $\theta_a$ for which the order fields $e^{-i\theta_a(\mathbf{x})G^a}$ are identical. The field $\theta_a(\mathbf{x})$ can therefore jump between such values without any cost in gradient energy. If we want to describe the statistical mechanics of NG-bosons, these jumps have to be included into the partition. This will be done with the help of vortex or defect gauge fields as we are now going to explain with the help of several examples.

The content of these lectures is largely contained in my textbook on this subject [1], except for discussions of some recent development.

## 2. Superfluid $^4$He

The simplest system containing NG-modes is superfluid $^4$He. They are describes by the phase factor of a complex order field

$$\psi(\mathbf{x}) = e^{i\theta(\mathbf{x})}\psi_0. \tag{9}$$

The line-like excitations are vortex lines.

### 2.1. Gradient Energy

In Landau's complex order field theory of superfluid $^4$He, the invariance under $U(1)$ transformation $\psi(\mathbf{x}) \to e^{i\alpha}\psi(\mathbf{x})$ is spontaneously broken. As a consequence, the phase fluctuations of $\psi(\mathbf{x})$ are massless. On account of its exponential nature, the order field (9) has the following important topological property: For every point $\mathbf{x}$ it is possible properly, let us introduce a $\delta$-function on a volume $V$

$$\delta(\mathbf{x}; V) = \int d^3\bar{x} \delta^{(3)}(\mathbf{x} - \bar{\mathbf{x}}). \tag{10}$$

Then the above topological property of the order field amounts to an invariance under the replacement

$$\theta(\mathbf{x}) \to \theta(\mathbf{x}) + 2\pi \delta(\mathbf{x}; V). \tag{11}$$

When expressing the gradient energy (8) in terms of $\partial\theta$, this invariance must be maintained.

Moreover, the freedom of jumping by $2\pi$ on a certain volume $V$ opens up the possibility of $\theta(\mathbf{x})$ jumping by $2\pi$ across some higher-dimensional manifold, for instance a surface $S$ in the three-dimensional superfluid. Due to the identity of the order field on both sides, the surface does not carry any energy. At the boundary $L$ of the surface, however, the gradient energy $\partial e^{i\theta(\mathbf{x})}$ does not behave smoothly. The boundary line carries an energy. It is a *vortex line*. To describe the energy of such a configuration we define a $\delta$-function on a surface in $D$ dimensions

$$\delta_{ij}(\mathbf{x}; S) \equiv \int ds dt \left[ \frac{\partial \bar{x}_j(s,t)}{\partial s} \frac{\partial \bar{x}_j(s,t)}{\partial t} - (st) \right] \delta^{(D)}(\mathbf{x} - \bar{x}(s,t)). \tag{12}$$

In three dimensions, we form

$$\delta_k(\mathbf{x}; S) \equiv \frac{1}{2} \epsilon_{ijk} \delta_{jk}(\mathbf{x}; S). \tag{13}$$

It is now easy to see that the euclidean action associated with the correct gradient energy of the NG-modes in the presence of a vortex line around $S$ must read

$$\mathcal{A}_{\text{NG}} = \frac{1}{2T} \int d^3x (\partial_i \theta - \theta_i^{\text{v}})^2 \tag{14}$$

where

$$\theta_i^{\text{v}}(\mathbf{x}) = 2\pi \delta_i(\mathbf{x}; S) \tag{15}$$

removes the singularities in $\partial_i \theta(\mathbf{x})$ arising at the physically irrelevant jumps of the field $\theta_i(\mathbf{x})$.

Since the surface $S$ is physically irrelevant, it can be deformed into any shape without changing the energy, as long as the boundary line stays the same. Indeed, under such a deformation $S \to S'$, the difference of the two surfaces encloses some volume $V$, and the $\delta$-function (13) changes as follows:

$$\delta_i(\mathbf{x}; S) \to \delta_i(\mathbf{x}; S') = \delta_i(\mathbf{x}; S) + \partial_i \delta(\mathbf{x}; V), \tag{16}$$

For the field $\theta_i^{\text{v}}(\mathbf{x})$, this implies

$$\theta_i^{\text{v}}(\mathbf{x}) \to \theta_i^{\text{v}}(\mathbf{x}) + \partial_i \Lambda_\delta(\mathbf{x}) \tag{17}$$

with the gauge function

$$\Lambda_\delta(\mathbf{x}) = 2\pi \delta(\mathbf{x}; V). \tag{18}$$

The energy (14) is invariant under this, if $\theta(\mathbf{x})$ is simultaneously transformed as

$$\theta(\mathbf{x}) \to \theta(\mathbf{x}) + \Lambda_\delta(\mathbf{x}) \tag{19}$$

The field $\theta_i^{\text{v}}$ is called the *vortex gauge field* of the superfluid and the combined transformation (17) and (19) is referred to as a *vortex gauge transformation*.

The subscript $\delta$ of the gauge function (18) emphasizes the fact that $\Lambda_\delta(\mathbf{x})$ is not an arbitrary function but a $\delta$-functions on an arbitrary volume.

The physical content of the vortex gauge field $\theta_i^v(\mathbf{x})$ appears when forming its curl. By Stokes' theorem we find

$$\epsilon_{ijk}\partial_j \delta_k(\mathbf{x}; S) = \delta_i(\mathbf{x}; L) \tag{20}$$

where

$$\delta_i(\mathbf{x}; L) = \int d\bar{x}_i \delta^{(3)}(\mathbf{x} - \bar{\mathbf{x}}) \tag{21}$$

is the $\delta$-function on the boundary line $L$. The curl of $\theta_i^v(\mathbf{x})$ is called the *vortex density*

$$\boldsymbol{\partial} \times \boldsymbol{\theta}^v(\mathbf{x}) = \mathbf{j}^v(\mathbf{x}) = 2\pi \boldsymbol{\delta}(\mathbf{x}; L) \tag{22}$$

As a consequence, the vortex density satisfies the obvious conservation law

$$\boldsymbol{\partial} \cdot \mathbf{j}^v(\mathbf{x}) = 0 \tag{23}$$

This states that vortex lines are closed.

The conservation law is a trivial consequence of $\mathbf{j}^v$ being the curl of $\boldsymbol{\theta}^v$. It is a Bianchi identity associated with the vortex gauge field structure.

The expression (14) may not yet be the complete energy of a vortex configuration. It is possible to add a gradient energy in the vortex gauge field. It introduces an extra *core energy* for the vortex line. The extended action is

$$\mathcal{A}_{\text{NG}} = \int d^3 x \left[ \frac{1}{2T} (\boldsymbol{\partial}\theta - \boldsymbol{\theta}^v)^2 + \frac{\epsilon_c}{2} (\boldsymbol{\partial} \times \boldsymbol{\theta}^v)^2 \right]. \tag{24}$$

The core energy term is a square of a $\delta$-function. It must be regularized by using a slightly smeared-out $\delta$-functions over some small length scale $a$. This length scale will be set equal to unity in the entire lecture. The regularizes last term is obviously proportional to the length of the lines.

### 2.3. The Partition Function

The partition function NG-modes and all possible vortex lines may be written as a functional integral

$$Z_{\text{NG}} = \int \mathcal{D}\theta \sum_{\{S\}} e^{-\mathcal{A}_{\text{NG}}}. \tag{25}$$

The measure $\int \mathcal{D}\theta$ is defined, as usual by discretizing the space into a simple cubic lattice of spacing an integration at each lattice point over all $\theta \in (-\infty, \infty)$, and taking the continuum limit $a \to 0$. The sum over all surface configurations is defined on the lattice ba setting

$$\theta_i^v(\mathbf{x}; S) \equiv 2\pi n_i(\mathbf{x}) \tag{26}$$

on each lattice point, where $n_i(\mathbf{x})$ is an integer-valued field, and by summing over all integer numbers $n_i(\mathbf{x})$

$$\sum_{\{S\}} = \sum_{\{n_i(\mathbf{x})\}}. \tag{27}$$

$$Z_{\text{latt}} = \left[ \prod_{\mathbf{x}} \int_{-\infty}^{\infty} d\theta(\mathbf{x}) \sum_{\{n_i(\mathbf{x})\}} \right]$$
$$\times e^{-\frac{1}{T} \int d^3x \left\{ \frac{1}{2}[\partial\theta(\mathbf{x}) - 2\pi\mathbf{n}(\mathbf{x})]^2 + \frac{\epsilon_c}{2}[\nabla \times \mathbf{n}(\mathbf{x})]^2 \right\}}. \tag{28}$$

which for $\epsilon_c = 0$ is the famous Villain approximation to the XY-model. The lattice spacing $a$ has been set equal to unity and the symbol $\nabla$ denotes the lattice derivative.

The vortex gauge fields requires a gauge fixing. On the lattice we can enforce the axial gauge

$$n_3(\mathbf{x}) = 0. \tag{29}$$

The Lorentz gauge $\nabla \cdot \mathbf{n} = 0$ cannot be chosen for integer-valued fields.

Similarly, the functional integral in the partition function (25) with the action (14) or (24) requires a gauge-fixing functional $\Phi[\boldsymbol{\theta}^{\text{v}}]$, for which we choose the $\delta$-functional

$$\Phi[\boldsymbol{\theta}^{\text{v}}] = \delta[\theta_3^{\text{v}}]. \tag{30}$$

Due to the sum over the vortex gauge fields $\boldsymbol{\theta}^{\text{v}}$, the partition function (25) describes superfluid $^4$He not only at zero temperature, where the NG-modes were identified, but at any not too large temperature. In particular, the phase transition is *included*. The vortex gauge field enriches the ensemble of fluctuating NG-modes in precisely the same way as the size of the order field $\psi$ does in a Landau description of the phase transition. In fact, it is easy to show that that near the transition, the partition function (25) can be transformes into a $|\psi|^4$-theory of the Landau type.

### 2.4. Interaction Energy between Vortices

Let us calculate the interaction energy between vortex lines. Omitting the core energy, we expand the action (14) into

$$\mathcal{A}_{\text{NG}} = \frac{1}{2T} \int d^3x \left\{ (\partial_i \theta)^2 + 2\theta \partial_i \theta_i^{\text{v}} + \theta_i^{\text{v}2} \right\}. \tag{31}$$

A quadratic completion leads to

$$\mathcal{A}_{\text{NG}} = \frac{1}{2T} \int d^3x \left\{ \left[ \partial_i \left( \theta + \frac{1}{-\partial^2} \partial_i \theta_i^{\text{v}} \right) \right]^2 + \left( \theta_i^{\text{v}2} - \partial_i \theta_i^{\text{v}} \frac{1}{-\partial^2} \partial_i \theta_i^{\text{v}} \right) \right\} \tag{32}$$

In the partition function, the $\theta$ field can be integrated out producing a factor $[\text{Det}(-\partial^2)]^{-1/2}$. The remaining action can be written as

$$\mathcal{A}_{\text{NG}} = \frac{1}{2T} \int d^3x (\partial \times \boldsymbol{\theta}^{\text{v}}) \frac{1}{-\partial^2} (\partial \times \boldsymbol{\theta}^{\text{v}})$$
$$= \frac{1}{2T} \int d^3x \, \mathbf{j}^{\text{v}} \frac{1}{-\partial^2} \mathbf{j}^{\text{v}}. \tag{33}$$

This has the form of the magnetic Biot-Savart energy for current loops, except for an opposite sign (parallel currents attract each other, parallel vortices repel each other).

The gauge freedom in the partition function (25) is destroyed if the original $U(1)$ symmetry is broken. This adds a mass term to the NG-modes, and the action (14) becomes

$$\mathcal{A}_{\mathrm{NG}} = \frac{1}{2T} \int d^3x \left\{ [\partial_i \theta(\mathbf{x}) - \theta_i^{\mathrm{v}}(\mathbf{x})]^2 + m^2 \theta(\mathbf{x})^2 \right\}. \tag{34}$$

As a consequence of the gauge dependence, the previously irrelevant surfaces become energetic. To see this we write the action as

$$\mathcal{A}_{\mathrm{NG}} = \frac{1}{2T} \int d^3x \left\{ [(\partial_i \theta)^2 + m^2 \theta^2] - \theta \partial_i \theta_i^{\mathrm{v}} + \theta_i^{\mathrm{v}2}] \right\}. \tag{35}$$

Integrating out the $\theta$-field in $Z$ leads to an interaction

$$\mathcal{A}_{\mathrm{int}} = \frac{1}{2T} \int d^3x \left[ \theta_i^{\mathrm{v}2} - \partial_i \theta_i \frac{1}{-\partial^2 + m^2} \partial_i \theta_i \right]. \tag{36}$$

This can be rewritten as

$$\mathcal{A}_{\mathrm{int}} = \frac{1}{2T} \int d^3 \left[ (\boldsymbol{\partial} \times \boldsymbol{\theta}^{\mathrm{v}}) \frac{1}{-\partial^2 + m^2} (\boldsymbol{\partial} \times \boldsymbol{\theta}^{\mathrm{v}}) + m^2 \theta_i^{\mathrm{v}} \frac{1}{-\partial^2 + m^2} \theta_i^{\mathrm{v}} \right]. \tag{37}$$

The first term is of the Biot-Savart type: antiparallel line elements attract each other with a finite-range Yukawa potential $e^{-mr}/4\pi r$. The second term describes a Yukawa interaction between the normal vector of the surface elements. It produces a physical surface of thickness $1/m$.

The same gauge-breaking mechanism can be used to construct a simple model of quark confinement which will be described in Section 4.

### 2.6. Gauge Field of Superflow

There exists an alternative representation of the partition function in which the vortex lines appear directly via their physical vortex density; the jumping surfaces $S$ being completely eliminated. This is possible by representing the NG-modes in terms of a new gauge field. It is canonically conjugate to the angular field $\theta$ and called generically the *stress gauge field*, which here is the *gauge field of superflow*. Recall that the momentum variable in an ordinary path integral

$$\int \mathcal{D}x \, e^{-\int dt \frac{\dot{x}^2}{2}} \tag{38}$$

is introduced by a quadratic completion, rewriting (38) as

$$\int \mathcal{D}x \mathcal{D}p \, e^{\int dt \left( ip\dot{x} - \frac{p^2}{2} \right)}. \tag{39}$$

By analogy, we rewrite the partition function (25) as

$$Z = \int \mathcal{D}\theta \int \mathcal{D}b_i \sum_{\{S\}} e^{-\mathcal{A}_{\mathrm{NG}}^c} \tag{40}$$

$$\mathcal{A}_{\mathrm{NG}}^{\mathrm{c}} = \int d^3x \left\{ \frac{1}{2} b_i^2(\mathbf{x}) - ib_i \left[ \partial_i \theta(\mathbf{x}) - \theta_i^{\mathrm{v}}(\mathbf{x}) \right] \right\}. \tag{41}$$

Note that if we go over to a Minkowski space formulation in which $x_0 = -ix^3$ plays the role of the time, the integral

$$\int \mathcal{D}b_0 e^{-ib_0(\mathbf{x})\partial_0 \theta} \tag{42}$$

creates on a discretized time axis a product of $\delta$-functions

$$\langle \theta_{n+1} | \theta_n \rangle \langle \theta_n | \theta_{n-1} \rangle \langle \theta_{n-1} | \theta_{n-2} \rangle \tag{43}$$

with

$$\langle \theta_n | \theta_{n-1} \rangle = \delta_n(\theta_n - \theta_{n-1}) \tag{44}$$

which can be interpreted as Dirac scalar products in the Hilbert space of the system. On this Hilbert space, there exists an operator $\hat{b}_i(\mathbf{x})$ whose zeroth component is given by

$$\hat{b}_0 = -i\partial_\theta \tag{45}$$

and satisfies the equal-time canonical communication rule

$$\left[ \hat{b}_0(\mathbf{x}_\perp, x_0), \theta(\mathbf{x}'_\perp, x_0) \right] = -i\delta^{(2)}(\mathbf{x}_\perp - \mathbf{x}'_\perp). \tag{46}$$

The charge associated with $\hat{b}_0(\mathbf{x})$,

$$\hat{Q}(\mathbf{x}_0) = \int d^2x \, \hat{b}_0(\mathbf{x}_\perp, x_0), \tag{47}$$

generates a constant shift in $\theta$:

$$e^{-i\hat{Q}(\mathbf{x}_0)} \theta(\mathbf{x}_\perp, x_0) e^{i\alpha\hat{Q}(\mathbf{x}_0)} = \theta(\mathbf{x}_\perp, x_0) + \alpha. \tag{48}$$

Thus it multiplies the original field $e^{i\theta(\mathbf{x})}$ by a phase factor $e^{i\alpha}$. The charge $\hat{Q}(\mathbf{x}_0)$ is the generator of the $U(1)$ symmetry transformation whose spontaneous symmetry breakdown is responsible for the NG-nature of the phase fluctuations $\theta(\mathbf{x})$. Since the original theory is invariant under the transformations $\psi \to e^{i\alpha}\psi$, the energy (41) does not depend on $\theta$ itself, only on $\partial_i \theta$. Thus, in the partition function, the field $\theta$ can be integrated out using the formula

$$\int \mathcal{D}\theta \, e^{if(\mathbf{x})\theta} = \delta[f(\mathbf{x})]. \tag{49}$$

This produces the conservation law

$$\partial_i b_i(\mathbf{x}) = 0, \tag{50}$$

implying that $\hat{Q}(\mathbf{x}_0)$ is a time independent charge and $e^{i\alpha\hat{Q}}$ a symmetry transformation.

If the energy does depend on $\theta$, the charge $\hat{Q}(\mathbf{x}_0)$ becomes $x_0$-dependent. However, it still generates the above $U(1)$ transformation.

---

[1]Similar canonical representations for the defect ensembles in a variety of physical systems in Ref. [2].

representation (41) contains more complicated functions of $b_i(\mathbf{x})$.

The conjugate variable to the phase angle of a complex field is the particle current, which is here the particle current in the superfluid condensate. Thus

$$j_i(\mathbf{x}) = b_i(\mathbf{x}) \tag{51}$$

can be interpreted as the *supercurrent*.

The dynamical conservation law $\partial_i b_i(\mathbf{x}) = 0$ can be ensured automatically as a Bianchi identity, if we represent $b_i(\mathbf{x})$ as a curl of a gauge field of superflow

$$b_i(\mathbf{x}) = \epsilon_{ijk} \partial_j a_k(\mathbf{x}). \tag{52}$$

Then the action (41) goes over into

$$\mathcal{A}'^{c}_{\mathrm{NG}} = \int d^3x \left[ \frac{T}{2}(\boldsymbol{\partial} \times \mathbf{a})^2 + i\mathbf{a} \cdot (\boldsymbol{\partial} \times \boldsymbol{\theta}^{\mathrm{v}}) + \frac{\epsilon_c}{2T}(\boldsymbol{\partial} \times \boldsymbol{\theta}^{\mathrm{v}})^2 \right] \tag{53}$$

The partition function becomes

$$Z = \int \mathcal{D}\mathbf{a} \sum_{\{S\}} e^{-\mathcal{A}^{c}_{\mathrm{NG}}} \tag{54}$$

The action is doubly gauge invariant, under the gauge transformations of superflow

$$a_i(\mathbf{x}) \to a_i(\mathbf{x}) + \partial_i \Lambda^{\mathrm{s}}(\mathbf{x}). \tag{55}$$

and under the vortex gauge transformation (17) and (18):

$$\theta_i^{\mathrm{v}}(\mathbf{x}) \to \theta_i^{\mathrm{v}}(\mathbf{x}) + \partial_i \Lambda_\delta(\mathbf{x}) \tag{56}$$

with the gauge functions

$$\Lambda_\delta(\mathbf{x}) = 2\pi \delta(\mathbf{x}; V). \tag{57}$$

With respect to the first gauge structure, the current conservation law $\partial_i b_i(\mathbf{x})$ is a Bianchi identity.

The integrations in the partition function (54) require a gauge fixing factor for both gauge fields. For the vortex gauge field, we use the axial factor (30), for the gauge field of superflow, any fixing will do, for instance the transverse one:

$$\Phi[\mathbf{a}] = \delta[\boldsymbol{\partial} \cdot \mathbf{a}]. \tag{58}$$

The action $\mathcal{A}'^{c}_{\mathrm{NG}}$ can also be written short as

$$\mathcal{A}'^{c}_{\mathrm{NG}} = \int d^3x \left[ \frac{T}{2}(\boldsymbol{\partial} \times \mathbf{a})^2 + i\mathbf{a} \cdot \mathbf{j}^{\mathrm{v}} + \frac{\epsilon_c}{2T}\mathbf{j}^{\mathrm{v}2} \right]. \tag{59}$$

In this expression, the jumping surfaces have disappeared and the action depends only on the vortex lines. The partition function may be rewritten as

$$Z = \int \mathcal{D}\mathbf{a} \sum_{\{L\}} e^{-\mathcal{A}'^{c}_{\mathrm{NG}}} \tag{60}$$

$$\Phi[\mathbf{j}^v] = \delta[\boldsymbol{\partial} \cdot \mathbf{j}^v] \tag{61}$$

which ensures the closedness of the vortex lines.

For a fixed set of lines $L$, the partition function (60) has a similar form as the magnetic partition function of a set of current lines. Around a vortex line, the field $\mathbf{b}(\mathbf{x}) = \boldsymbol{\partial} \times \mathbf{a}(\mathbf{x})$ looks precisely like a magnetic field $\mathbf{B}(\mathbf{x}) = \boldsymbol{\partial} \times \mathbf{A}(\mathbf{x})$ around a current line. Thus, integrating out the vector potential $\mathbf{a}$ yields once more the same Biot-Savart interaction energies as obtained before in another way in Eq. (33).

Note that if the energy contains an explicit dependence like the mass term of the last section, there exists no $\mathbf{a}$-field representation of the NG-modes.

### 2.7. Disorder Field Theory

The sum over all vortex line configurations in the partition function (60) can be done separately, defining an $\mathbf{a}$-dependent vortex partition function

$$\begin{aligned} Z^v[\mathbf{a}] &= \sum_{\{S\}} \delta[\theta_3^v] \exp\left\{-\int d^3x \left[\frac{\epsilon_c}{2T}(\boldsymbol{\partial} \times \boldsymbol{\theta}^v)^2 + i\mathbf{a} \cdot (\boldsymbol{\partial} \times \boldsymbol{\theta}^v)\right]\right\} \\ &= \sum_{\{L\}} \delta[\boldsymbol{\partial} \cdot \mathbf{j}^v] \exp\left\{-\int d^3x \left[\frac{\epsilon_c}{2T}\mathbf{j}^{v2} + i\mathbf{a} \cdot \mathbf{j}^v\right]\right\}. \end{aligned} \tag{62}$$

This can be viewed as a particular representation of an entirely different set of NG-modes, the dual NG-modes. To see this we make use of an important relation:

$$\sum_{\{L\}} e^{i \int d_x \theta_i^v(\mathbf{x}) \delta_i(\mathbf{x})} = \sum_{\{\tilde{L}\}} \delta\left[f_i(\mathbf{x}) - \delta_i(\tilde{L})\right]. \tag{63}$$

This can easily be proved by going on a lattice and using the Poisson formula

$$\sum_n e^{i2\pi n \mu} = \sum_m \delta(\mu - m). \tag{64}$$

It is therefore possible to rewrite the partition function (62) as follows

$$Z^v[\mathbf{a}] = \sum_{\{\tilde{L}\}} \int \mathcal{D}b_i \int \mathcal{D}\tilde{\theta} \exp\left\{-\int d^3x \left[\frac{\epsilon_c}{2T}j_i^{v2} + ia_i b_i + \frac{i}{2\pi}j_i^v\left(\partial_i\tilde{\theta} - \tilde{\theta}_i^v\right)\right]\right\}. \tag{65}$$

Integrating out the $b_i(\mathbf{x})$-field gives

$$Z^v[\mathbf{a}] = \sum_{\{\tilde{L}\}} \int \mathcal{D}\tilde{\theta} \exp\left[-\frac{1}{8\pi^2}\frac{T}{\epsilon_c}\int d^3x(\partial_i\tilde{\theta} - \tilde{\theta}_i^v - 2\pi a_i)^2\right]. \tag{66}$$

This is the NG-theory of another $U(1)$-invariant $|\phi^4|$-theory, the above-mentioned disorder field theory. It differs from the initial $|\psi|^4$ theory by the presence of an external vortex gauge field $\mathbf{a}$ in the gradient term coupled via a covariant derivative. The disorder field partition function is

$$Z^v[\mathbf{a}] = \int \mathcal{D}\psi \mathcal{D}\psi^* \exp\left\{-\int d^3x \left[\left|(\partial_i - 2\pi a_i)\varphi\right|^2 + m^2|\varphi|^2 + \frac{g}{4}|\varphi|^4\right]\right\}. \tag{67}$$

frozen out. The functional integral over $\tilde{\theta}$ removes the longitudinal part of $a_i$, and we remain with

$$Z^{\mathrm{v}}[\mathbf{a}] = \exp\left(-T\frac{m_a^2}{2}\int d^3x\, \mathbf{a}_T^2\right) \tag{68}$$

where

$$m_a^2 = \frac{1}{\epsilon_c}. \tag{69}$$

The exponent in (68) is a mass term for the transverse projection of the gauge field of superflow

$$a_i^T \equiv \left(\delta_{ij} - \partial_i\partial_j/\partial^2\right)a_j. \tag{70}$$

In the complex disorder field theory (67), the mass term $m^2$ is negative and the disorder field acquires a nonzero expectation value $\psi_0 = \sqrt{-m^2/g}$. This produces again a mass term (68).

The mass term (68) shows that in the high-temperature phase, the singular nature of the vortex gauge field $\boldsymbol{\theta}^{\mathrm{v}}$ is lost and $\boldsymbol{\theta}^{\mathrm{v}}$ can be treated as an ordinary fluctuating transverse field. The sum $\sum_{\{\tilde{L}\}}$ turns into a functional integral over an ordinary gauge field $\int \mathcal{D}\boldsymbol{\theta}^{\mathrm{v}}$. Indeed, when doing this integral in the partition function out (62), we find that the vortex density has the simple correlation function

$$<j_i^{\mathrm{v}}(\mathbf{x})j_j^{\mathrm{v}}(\mathbf{x}')> = \frac{T}{\epsilon_c}(\delta_{ij} - \partial_i\partial_j/\partial^2)\delta^{(3)}(\mathbf{x}-\mathbf{x}'), \tag{71}$$

which produces directly the mass term (68).

This is to be contrasted with the contributions to the $\mathbf{a}$-field action coming from $Z^v[\mathbf{a}]$ of Eq. (62) at low temperatures. There the vortex lines form small loops, and these can be shown to give[2]

$$Z^{\mathrm{v}}[\mathbf{a}] \sim \exp\left\{-\frac{1}{2}(\partial\times\mathbf{a})f(-i\partial)(\partial\times\mathbf{a})\right\}. \tag{72}$$

Such a contribution changes only the dispersion of the gauge fields of superflow. Infinitely long vortex lines are necessary to change the mass. These are present at high temperatures where the correlation function of the vortex densities is (71).

With the help of the disorder partition function $Z^{\mathrm{v}}[\mathbf{a}]$, the partition function (25) of the NG-modes associated with the action (24) can be replaced by another completely equivalent one:

$$\tilde{Z}_{\mathrm{NG}} = \int \mathcal{D}\mathbf{a}\sum_{\{\tilde{L}\}}\int \mathcal{D}\tilde{\theta}\, e^{-\tilde{\mathcal{A}}_{\mathrm{NG}}} \tag{73}$$

with the action

$$\tilde{\mathcal{A}}_{\mathrm{NG}} = T\int d^3x\left\{\frac{1}{2}(\partial\times\mathbf{a})^2 + \frac{m_a^2}{8\pi^2}\left(\partial\tilde{\theta} - \tilde{\boldsymbol{\theta}}^{\mathrm{v}} - 2\pi\mathbf{a}\right)^2\right\}. \tag{74}$$

This action is invariant under the following two types of gauge transformation. First, there is invariance under the gauge transformations of superflow (55), if it is accompanied by a compensating transformation of the angular field $\tilde{\theta}$:

$$a_i(\mathbf{x}) \to a_i(\mathbf{x}) + \partial_i\Lambda(\mathbf{x}), \qquad \tilde{\theta}(\mathbf{x}) \to \tilde{\theta}(\mathbf{x}) + 2\pi\Lambda(\mathbf{x}). \tag{75}$$

---

[2] See Ref. [1], Vol. I, Part 2, Section 9.7 and Ref. [3].

field, the analogs of Eqs. (17)–(19):

$$\tilde{\theta}_i^v(\mathbf{x}) \to \tilde{\theta}_i^v(\mathbf{x}) + \partial_i \tilde{\Lambda}_\delta(\mathbf{x}), \qquad \tilde{\theta}(\mathbf{x}) \to \tilde{\theta}(\mathbf{x}) + \tilde{\Lambda}_\delta(\mathbf{x}), \tag{76}$$

with gauge functions

$$\tilde{\Lambda}_\delta(\mathbf{x}) = 2\pi\delta(\mathbf{x}; \tilde{V}). \tag{77}$$

At high temperatures, the vortex lines in $\boldsymbol{\theta}^v$ are frozen out and the action (74) shows again the mass term (68).

The mass term implies that at high temperatures, the gauge field of superflow possesses a finite range. At some critical temperature superfluidity has been destroyed. This is an analog of the Meissner effect within the disorder theory [4]. Note that in the absence of the gauge field of superflow $\mathbf{a}$, the field $\tilde{\theta}$ would be of long range, i.e., massless. The gauge field of superflow absorbs the massless mode and the system has only short-range excitations. More precisely, it can be shown that any correlation function involving local gauge-invariant observable quantities must be of short range in high-temperature phase.

Take, for instance, the local gauge-invariant current operator of the disorder field

$$\mathbf{j}^s \equiv \boldsymbol{\partial}\tilde{\theta} - \mathbf{a} \tag{78}$$

In the gauge $\tilde{\theta} \equiv 0$, we easily find

$$\langle j_i^s(\mathbf{x_1}) j_j^s(\mathbf{x_2}) \rangle \propto \int d^3 x\, e^{i\mathbf{k}(\mathbf{x_1}-\mathbf{x_2})} \frac{\delta_{ij} - k_i k_j / m_a^2}{\mathbf{k}^2 + m_a^2}. \tag{79}$$

This has *no* zero-mass pole.

If we remove the locality restriction, it is easy to find a filed which has long-range correlations in the superconducting phase. If there were no magnetic interactions, the exponential filed $e^{i\tilde{\theta}}$ would provides a order parameter of the system (which would then be an ordinary superfluid of Cooper pairs):

$$\tilde{\mathcal{O}} = \langle e^{i\tilde{\theta}} \rangle. \tag{80}$$

In the presence of the vector potential $\mathbf{A}$, this is no longer true since $e^{i\tilde{\theta}}$ it is not gauge invariant. Its expectation value vanishes identically on account of Elitzur's theorem [5]. There exists, however, a simple gauge-invariant modification of this used first by Dirac [6, 7]. Let

$$Q(\mathbf{x}') \equiv \delta^{(3)}(\mathbf{x}' - \mathbf{x}) \tag{81}$$

be a charge density localized at the point $\mathbf{x}$. Then the disorder field

$$\tilde{O}(\mathbf{x}) \equiv e_{\mathrm{gi}}^{i\tilde{\theta}(\mathbf{x})} \equiv \exp\left[i \int d^3 x \left(\tilde{\theta} + \frac{\mathbf{A}\cdot\boldsymbol{\partial}}{\boldsymbol{\partial}^2}\right) Q\right] \tag{82}$$

is gauge invariant and the expectation value

$$\tilde{\mathcal{O}} = \langle e_{\mathrm{gi}}^{i\tilde{\theta}(\mathbf{x})} \rangle \tag{83}$$

is nonzero.

Note that apart from being invariant under magnetic gauge transformations

$$A_i(\mathbf{x}) \to A_i(\mathbf{x}) + \partial_i \Lambda(\mathbf{x}), \tag{84}$$

At low temperatures where the vortex lines can be ignored, it can easily be shown to be nonzero. We calculate the Green function associated with two fields (82):

$$G_{\tilde{\mathcal{O}}}(\mathbf{x}_1, \mathbf{x}_2) \equiv \langle e^{i\tilde{\theta}(\mathbf{x}_1)}_{\mathrm{gi}} e^{-i\tilde{\theta}(\mathbf{x}_2)}_{\mathrm{gi}} \rangle. \tag{85}$$

In the transverse gauge, we can easily integrate out the $\tilde{\theta}$-fluctuations in (73) and find the same correlation function as in (111)

$$G_{\tilde{\mathcal{O}}}(\mathbf{x}_1, \mathbf{x}_2) \xrightarrow[|\mathbf{x}_1-\mathbf{x}_2|\to\infty]{} \text{const.} \tag{86}$$

Hence

$$\tilde{\mathcal{O}} = \langle e^{\tilde{\theta}(\mathbf{x})_{\mathrm{gi}}} \rangle = \text{const.} \tag{87}$$

This is nonzero, just as the order parameter of the superfluid in the low temperature phase.

### 2.8. Disorder Theory of Superconductor

A superconductor is described by the Ginzburg-Landau action

$$\mathcal{A}_{\mathrm{GL}} = \frac{1}{\tilde{T}} \int d^3x \left\{ |(\eth - i\mathbf{A})\varphi|^2 + m^2|\varphi|^2 + \frac{g}{2}|\varphi|^4 + \frac{1}{2}(\eth \times \mathbf{A})^2 \right\}. \tag{88}$$

We have set the charge of the Cooper pairs equal to unity, for simplicity. In the superconducting phase, $m^2$ is negative. Going to the *London limit*, the field $\varphi$ is approximated by

$$\varphi = e^{i\tilde{\theta}(\mathbf{x})} \varphi_0 \tag{89}$$

with a constant real $\varphi_0 = \sqrt{-m^2/g}$, and the action becomes

$$\mathcal{A}_{\mathrm{LL}} = \frac{1}{\tilde{T}} \int d^3x \left[ \frac{m_A^2}{2} \left( \eth\tilde{\theta} - \tilde{\boldsymbol{\theta}}^{\mathrm{v}} - \mathbf{A} \right)^2 + \frac{1}{2}(\eth \times \mathbf{A})^2 \right] \tag{90}$$

where we have set $\varphi_0^2 = m_A^2/2e^2$. The partition function reads

$$Z_{\mathrm{SC}} = \int \mathcal{D}\tilde{\theta} \int \mathcal{D}\mathbf{A} \sum_{\{\tilde{L}\}} e^{-\mathcal{A}_{\mathrm{LL}}}. \tag{91}$$

The action (90) has the same form as the action in the disorder representation (74) of superfluid $^4$He. The role of the gauge field of superflow is now played by the vector potential $\mathbf{A}$ of magnetism. The action has the following two types of gauge invariances: the magnetic ones

$$A_i(\mathbf{x}) \to A_i(\mathbf{x}) + \partial_i \Lambda(\mathbf{x}), \qquad \tilde{\theta}(\mathbf{x}) \to \tilde{\theta}(\mathbf{x}) + \Lambda(\mathbf{x}), \tag{92}$$

and the vortex gauge transformations

$$\tilde{\theta}_i^{\mathrm{v}}(\mathbf{x}) \to \tilde{\theta}_i^{\mathrm{v}}(\mathbf{x}) + \partial_i \tilde{\Lambda}_\delta(\mathbf{x}), \qquad \tilde{\theta}(\mathbf{x}) \to \tilde{\theta}(\mathbf{x}) + \tilde{\Lambda}_\delta(\mathbf{x}), \tag{93}$$

with gauge functions

$$\tilde{\Lambda}_\delta(\mathbf{x}) = 2\pi \delta(\mathbf{x}; \tilde{V}). \tag{94}$$

function (91) describes the statisctical behavior of the superconductor not only at zero temperature, where the action (90) was derived, but at all not too large temperatures. The fluctuating vortex gauge field $\boldsymbol{\theta}^{\rm v}$ ensures the validity through the phase transition.

We shall now derive the disorder representation of this partition function in which the vortex lines of the superconductor play a central role in describing the phase transition.

At low temperatures, the vortices are frozen, and the $\tilde{\theta}$-fluctuations in the partition function can be integrated out. This reduces the action to

$$\mathcal{A}_{\rm LL} \sim \frac{m_A^2}{2\tilde{T}} \int d^3x\, \mathbf{A}_T^2, \tag{95}$$

i.e., to a simple transverse mass term for the vector potential $\mathbf{A}$. This is the famous Meissner effect in a superconductor, which limits the range of a magnetic field to a finite *penetration depth* $\lambda = 1/m_A$. The effect is completely analogous to the one observed previously in the disorder description of the superfluid where the superfluid acquired a finite range in the normal phase.

To derive the disorder theory of the partition function (91), we supplement the action by a core energy of the vortex lines

$$\mathcal{A}_{\rm core} = \frac{\tilde{\epsilon}_c}{2\tilde{T}} \int d^3x\, (\boldsymbol{\partial} \times \tilde{\boldsymbol{\theta}}^{\rm v})^2. \tag{96}$$

As in the partition function (40), an auxiliary $\tilde{b}_i$ field can be introduced to bring the action (90) to the form

$$\tilde{\mathcal{A}}_{\rm LL} = \int d^3x \left[ \frac{\tilde{T}}{2m_A^2}\tilde{\mathbf{b}}^2 + i\tilde{\mathbf{b}} \left( \boldsymbol{\partial}\tilde{\theta} - \tilde{\boldsymbol{\theta}}^{\rm v} - \mathbf{A} \right) + \frac{1}{2\tilde{T}}(\boldsymbol{\partial} \times \mathbf{A})^2 + \frac{\tilde{\epsilon}_c}{2\tilde{T}}(\boldsymbol{\partial} \times \tilde{\boldsymbol{\theta}}^{\rm v})^2. \right] \tag{97}$$

By integrating out the $\tilde{\theta}$-fields, we obtain the conservation law

$$\boldsymbol{\partial} \cdot \tilde{\mathbf{b}} = 0 \tag{98}$$

which we fulfill by introducing the gauge field of superflow in the superconductor

$$\tilde{\mathbf{b}} = \boldsymbol{\partial} \times \tilde{\mathbf{a}}. \tag{99}$$

In this way, we find the action

$$\tilde{\mathcal{A}}_{\rm LL} = \int d^3x \left[ \frac{\tilde{T}}{2m_A^2}(\boldsymbol{\partial} \times \tilde{\mathbf{a}})^2 - i\tilde{\mathbf{a}} \cdot (\boldsymbol{\partial} \times \mathbf{A}) + \frac{1}{2\tilde{T}}(\boldsymbol{\partial} \times \mathbf{A})^2 - i\tilde{\mathbf{a}} \cdot \tilde{\mathbf{j}}^{\rm v} + \frac{\tilde{\epsilon}_c}{2\tilde{T}}\tilde{\mathbf{j}}^{\rm v 2} \right]. \tag{100}$$

where

$$\tilde{\mathbf{j}}^{\rm v} = \boldsymbol{\partial} \times \tilde{\boldsymbol{\theta}}^{\rm v}. \tag{101}$$

At low temperatures $\tilde{T}$ where vortex lines are absent, the $\tilde{\mathbf{a}}$-field can be integrated out and yields once more the transverse mass term (95).

At high temperatures $\tilde{T}$, the vortex lines are prolific and the vector vortex gauge field $\boldsymbol{\theta}^{\rm v}$ can be integrated out like an ordinary field. With the help of the correlation function (71), we obtain the transverse mass term

$$\frac{\tilde{T}}{2m_A^2} \int d^3x\, \frac{m_a^2}{2}\tilde{\mathbf{a}}_T^2 \tag{102}$$

$$m_a^2 = 1/m_A^2 \tilde{\epsilon}_c. \tag{103}$$

When integrating out the massive $\tilde{\mathbf{a}}$-field, no mass term is produced for the vector potential $\mathbf{A}$. Instead, the vector potential acquires the effective action

$$\mathcal{A}_{\mathbf{A}} = \frac{1}{2\tilde{T}} \int d^3x\, \partial \times \mathbf{A} \left(1 + \frac{m_A^2}{-\partial^2 + m_a^2}\right) \partial \times \mathbf{A}. \tag{104}$$

Expanding the denominator in powers of $-\partial^2$ we see that only gradient energies appear. Thus, apart from a modification of the dispersion, the vector potential is of long range giving Coulomb-like forces at large distances.

For the low-temperature mass value $m_a^2 = 0$, the second term reduces again a transverse mass term (95) associated with the Meissner effect.

We can represent the fluctuating vortices in the superconductor by a disorder field theory in the same way as we did for the vortices in the superfluid, i.e., we repeat the transformations from (62) to (66). The angular field variable of disorder will be called $\theta$, the vortex lines in the disorder theory $\boldsymbol{\theta}^{\mathrm{v}}$. The disorder action reads

$$\tilde{\mathcal{A}}_{\mathrm{LL}}^{\mathrm{D}} = \int d^2x \left[ \frac{\tilde{T}}{2m_A^2}(\partial \times \tilde{\mathbf{a}})^2 - i\tilde{\mathbf{a}} \cdot (\partial \times \mathbf{A}) + \frac{1}{2\tilde{T}}(\partial \times \mathbf{A})^2 + \frac{\tilde{T}}{8\pi^2 m_A^2} m_a^2 (\partial \theta - \boldsymbol{\theta}^{\mathrm{v}} - 2\pi \tilde{\mathbf{a}})^2 \right]. \tag{105}$$

Near the phase transition, this is equivalent to a disorder field action

$$\tilde{\mathcal{A}}_{\mathrm{LL}}^{\mathrm{D}} \sim \int d^2x \left\{ \frac{\tilde{T}}{2m_A^2}(\partial \times \tilde{\mathbf{a}})^2 - i\tilde{\mathbf{a}} \cdot (\partial \times \mathbf{A}) + \frac{1}{2\tilde{T}}(\partial \times \mathbf{A})^2 \right.$$
$$\left. + |(\partial - 2\pi\tilde{\mathbf{a}})\psi|^2 + m^2|\psi|^2 + \frac{g}{2}|\psi|^4 \right\}. \tag{106}$$

The loop diagrams of the disorder field $\psi$ are the pictures of the vortex lines.

Note that on the average, $i\tilde{\mathbf{a}}$ is equal to the magnetic field $\partial \times \mathbf{A}$.

The disorder fiels theory for the superconductor has recently led to a determination of the critical properties of the superconductive phase transition in a $4 - \epsilon$ expansion [8]. Within the Ginzburg-Landau theory, this is impossible since the renormalization group gives no infrared stable fixed point at all.

### 2.9. Order versus Disorder

*Superfluid* $^4He$

The order parameter $\mathcal{O}$ of the original theory (14) of (24) is defined as the expectation value of the phase of the complex order field $O(\mathbf{x}) = e^{i\theta(\mathbf{x})}$:

$$\mathcal{O} \equiv \langle O(\mathbf{x}) \rangle = \langle e^{i\theta(\mathbf{x})} \rangle. \tag{107}$$

Its behavior can best be studied by considering the correlation function of two order fields $O(\mathbf{x})$:

$$G_O(\mathbf{x}, \mathbf{x}') = \langle e^{i\theta(\mathbf{x})} e^{-i\theta(\mathbf{x}')} \rangle. \tag{108}$$

In the low temperature phase, where vortices are rare the $\theta(\mathbf{x})$-field fluctuates almost harmonically. By Wick's theorem, we can then approximate

$$G_O(\mathbf{x}_1, \mathbf{x}_2) \approx e^{-\frac{1}{2}\langle[\theta(\mathbf{x}_1) - \theta(\mathbf{x}_2)]^2\rangle}. \tag{109}$$

$$\langle\theta(\mathbf{x}_1)\theta(\mathbf{x}_2)\rangle \approx T\int\frac{d^3k}{(2\pi)^3}e^{i\mathbf{k}(\mathbf{x}_1-\mathbf{x}_2)}\frac{1}{\mathbf{k}^2}=T\frac{1}{4\pi|\mathbf{x}_1-\mathbf{x}_2|} \tag{110}$$

goes to zero for $r\to\infty$. After a renormalization of the fields, we see that

$$G_O(\mathbf{x}_1,\mathbf{x}_2)\xrightarrow[|\mathbf{x}_1-\mathbf{x}_2|\to\infty]{}\text{const.} \tag{111}$$

By the cluster properties of correlation functions,

$$\langle O(\mathbf{x}_1)O(\mathbf{x}_2)\rangle\xrightarrow[|\mathbf{x}_1-\mathbf{x}_2|\to\infty]{}\langle O(\mathbf{x}_1)\rangle\langle O(\mathbf{x}_2)\rangle, \tag{112}$$

this implies that the order parameter is a constant:

$$\mathcal{O}=\langle O(\mathbf{x})\rangle=\text{const.} \tag{113}$$

Let us calculate the order parameter in the high temperature phase. To find the correlation function $G_O(\mathbf{x}_1,\mathbf{x}_2)$, we insert into the partition function a term

$$e^{-iq(\mathbf{x})\theta(\mathbf{x})} \tag{114}$$

with two point sources of opposite charges at the points $\mathbf{x}_1$ and $\mathbf{x}_2$:

$$q(\mathbf{x})=\delta^{(3)}(\mathbf{x}-\mathbf{x}_1)-\delta^{(3)}(\mathbf{x}-\mathbf{x}_2). \tag{115}$$

When going to the canonical representation (41), we obtain an action

$$\mathcal{A}=\int d^3x\left[\frac{T}{2}b_i^2-ib_i\left(\partial_i\theta-\theta_i^{\text{v}}\right)+iq(\mathbf{x})\theta(\mathbf{x})\right]. \tag{116}$$

Integrating out the $\theta$ field in the partition function gives the constraint

$$\partial_ib_i(\mathbf{x})=-q(\mathbf{x}). \tag{117}$$

In the magnetic-field analogy, the point charges in $q(\mathbf{x})$ correspond to a magnetic monopole and an anti-monopole.

The constraint is solved by introducing a singular field

$$b_i^{\text{m}}(\mathbf{x})=-\delta_i(\mathbf{x};\bar{L}) \tag{118}$$

where the line $\bar{L}$ runs from $\mathbf{x}_1$ to $\mathbf{x}_2$ along an arbitrary path. This field satisfies

$$\partial_ib_i^{\text{m}}(\mathbf{x})=-q(\mathbf{x}). \tag{119}$$

We can therefore write

$$\mathbf{b}(\mathbf{x})=\boldsymbol{\partial}\times\mathbf{a}(\mathbf{x})-\mathbf{b}^{\text{m}}(\mathbf{x}), \tag{120}$$

and the action (116) can be replaced by

$$\mathcal{A}=\int d^3x\left[\frac{T}{2}\left(\boldsymbol{\partial}\times\mathbf{a}-\mathbf{b}^{\text{m}}\right)^2-i\mathbf{a}\cdot\mathbf{j}^{\text{v}}+\frac{\epsilon_c}{2T}\mathbf{j}_c^{\text{v}2}\right]. \tag{121}$$

fixed ends results in a transformation

$$b_i^m(\mathbf{x}) \to b_i^m(\mathbf{x}) + \epsilon_{ijk}\partial_j \delta_k(\mathbf{x}; S), \tag{122}$$

where $S$ is the surface over which the line $\bar{L}$ has swept. The action (121) remains invariant if the gauge field of superflow is simultaneously transformes as

$$a_i(\mathbf{x}) \to a_i(\mathbf{x}) + \delta_i(\mathbf{x}; S). \tag{123}$$

The pure phase order field $e^{i\theta(\mathbf{x})}$ corresponds to a gauge field

$$b_i^m(\mathbf{x}) = \delta_i(\mathbf{x}; \bar{L}_1) \tag{124}$$

with a line $\bar{L}_1$ starting at $\mathbf{x}$ and running to infinity, i.e., it is represented in the path integral (73) by the integrand

$$e^{i\theta(\mathbf{x})} \triangleq e^{-T \int d^3x \left\{\frac{1}{2}\mathbf{b}^m(\mathbf{x})^2 - \mathbf{b}^m(\mathbf{x}) \cdot [\partial \times \mathbf{a}(\mathbf{x})]\right\}}. \tag{125}$$

When integrating out the gauge field of superflow $\mathbf{a}$ at low temperatures where the vortex lines are frozen out, the action the partition function contains $\mathbf{b}^m$ in the form of a factor

$$e^{-\frac{T}{2} \int d^3x \left[\mathbf{b}^{m2} - (\partial \times \mathbf{b}^m)\frac{1}{-\partial^2}(\partial \times \mathbf{b}^m)\right]} = e^{-\frac{T}{2} \int d^3x\, \partial \cdot \mathbf{b}^m \frac{1}{-\partial^2} \partial \cdot \mathbf{b}^m}. \tag{126}$$

From this we obtain the correlation function

$$G_O(\mathbf{x}_1, \mathbf{x}_2) = e^{-\frac{T}{2} \int d^3x q \frac{1}{-\partial^2} q}. \tag{127}$$

Inserting (119), this is precisely the correlation function (109) with (110), which goes to a constant for $|\mathbf{x}_1 - \mathbf{x}_2'| \to \infty$ as in (111) implying a nonzero order parameter.

In the disorder version $\tilde{Z}_{\mathrm{NG}}$ of the partition function $Z_{\mathrm{NG}}$, it is easy to calculate the correlation function also in the high-temperature phase. There, the $\mathbf{a}$-field contains the transverse mass term (68) and the situation is very similar to that in Section 2.5. The line connecting the points $\mathbf{x}_1$ and $\mathbf{x}_2$ becomes physical and its energy grows linearly with the length of the line. Thus, the correlation function vanishes

$$G_O(\mathbf{x}_1, \mathbf{x}_2) \sim e^{-\mathrm{const} \cdot |\mathbf{x}_1 - \mathbf{x}_2|} \xrightarrow[|\mathbf{x}_1 - \mathbf{x}_2| \to \infty]{} 0. \tag{128}$$

By the cluster properties (112) of correlation functions, this shows that the order parameter vanishes.

Note that precisely the same exponential falloff is found within Landau's complex order field theory:

$$<\psi(\mathbf{x}_1)\psi(\mathbf{x}_2)> \propto \int \frac{d^3k}{(2\pi)^3} e^{i\mathbf{k}(\mathbf{x}_1-\mathbf{x}_2)} \frac{1}{\mathbf{k}^2 + m^2} = \frac{1}{4\pi} \frac{e^{-m|\mathbf{x}_1-\mathbf{x}_2|}}{|\mathbf{x}_1 - \mathbf{x}_2|}. \tag{129}$$

The mechanism by which the finite range arises is, however, quite different. The size fluctuations of the order field play an essential role. In the partition function (25), their end of Section 2.3.

---

[3]See Ref. [1], Vol. I, Part 2, Chapter 10.

*b) Superconductor*

The order parameter of the superfluid supplies directly a disorder parameter of the superconductor [9]. Inserting the monopole gauge filed (124) into the actions (100), (105) and (105), we obtain the disorder field analogous to (125):

$$\tilde{D}(\mathbf{x}) \equiv e^{-\frac{1}{\tilde{T}} \int d^3x \left\{\frac{1}{2}\mathbf{b^m}(\mathbf{x})^2 - \mathbf{b^m}(\mathbf{x}) \cdot [\partial \times \mathbf{A}(\mathbf{x})]\right\}}. \tag{130}$$

The action for the correlation function of a monopole antimonopole pair is

$$\mathcal{A} = \int d^3x \left[ \frac{\tilde{T}}{2m_A^2} (\partial \times \tilde{\mathbf{a}})^2 - i\tilde{\mathbf{a}} \cdot (\partial \times \mathbf{A}) + \frac{1}{2\tilde{T}} (\partial \times \mathbf{A} - \mathbf{b^m})^2 - i\mathbf{a} \cdot \tilde{\mathbf{j}}^{\mathrm{v}} + \frac{\tilde{\epsilon}_c}{2\tilde{T}} \tilde{\mathbf{j}}_c^{\mathrm{v}2} \right]. \tag{131}$$

with $\mathbf{b^m}$ of Eq. (118). At low temperature $\tilde{T}$ of the superconductor, where the vortex lines of the superconductor are frozen out and the integral over the $\tilde{\mathbf{a}}$-filed yields a mass term for the vector potential $\mathbf{A}$, this becomes

$$\mathcal{A} = \int d^3x \left\{ \frac{1}{2\tilde{T}} \left[ (\partial \times \mathbf{A} - \mathbf{b^m})^2 + \frac{m_A^2}{2} \mathbf{A}_T^2 \right] - i\mathbf{a} \cdot \tilde{\mathbf{j}}^{\mathrm{v}} + \frac{\tilde{\epsilon}_c}{2\tilde{T}} \tilde{\mathbf{j}}_c^{\mathrm{v}2} \right\}. \tag{132}$$

Just as in (128), the line connecting the endpoints acquires an energy and the correlation function has the property

$$G_{\tilde{D}}(\mathbf{x}_1, \mathbf{x}_2) \sim e^{-\mathrm{const} \cdot |\mathbf{x}_1 - \mathbf{x}_2|} \xrightarrow[|\mathbf{x}_1 - \mathbf{x}_2| \to \infty]{} 0. \tag{133}$$

implying a vanishing disorder parameter:

$$\tilde{\mathcal{D}} = \langle \tilde{D}(\mathbf{x}) \rangle = 0. \tag{134}$$

In the normal phase of the superconductor, the monopole gauge field enters the action (104) which becomes

$$\mathcal{A}'_{\mathbf{A}} = \frac{1}{2\tilde{T}} \int d^3x \, (\partial \times \mathbf{A} - \mathbf{b^m}) \left( 1 + \frac{m_A^2}{-\partial^2 + m_a^2} \right) (\partial \times \mathbf{A} - \mathbf{b^m}). \tag{135}$$

As in (127), we obtain for a large distance between the monopole and the antimonopole a Coulomb interaction implying that the disorder parameter is a constant:

$$\tilde{\mathcal{D}} = \langle \tilde{D}(\mathbf{x}) \rangle = \mathrm{const}. \tag{136}$$

## 2.10. Order of Superconducting Phase Transition—Tricritical Point

In the superconductive phase, there are only a few vortex lines and the field $\psi$ fluctuates around zero. The mass term is $m^2 > 0$. The vector potential $\mathbf{A}$ may be integrated out and leads to an action for the **a**-field

$$\mathcal{A}_{\mathbf{a}} = \int d^3x \left[ \frac{1}{2m_A^2} (\partial \times \mathbf{a})^2 + \frac{1}{2}\mathbf{a}_T^2 + \mathbf{a}_T^2 \frac{c}{2} |\psi|^2 \right] \tag{137}$$

between **a** and $\psi$ simple. This yields also a coupling term $\mathbf{a}_L^2|\psi|^2$ involving the longitudinal components of the field **a**. When integrating out the longitudinal part, the functional determinant is canceled by the gauge fixing factor of the gauge $\psi$=real.

The **a**-field can now be integrated out; for a constant $\psi$ they yield an effective action

$$\mathcal{A} = \frac{1}{2}\mathrm{Tr}\log\left(-\partial^2 + m_A^2 + m_A^2 c|\psi|^2\right). \tag{138}$$

When expanded in powers of $|\psi|^2$, this produces a term

$$-c^2 m_A^4 \int d^3x \int \frac{d^3k}{(2\pi)^3}\frac{1}{(k^2+m_A^2)^2}|\psi|^4 \propto -m_A^3 \int d^3x|\psi|^4. \tag{139}$$

This term lowers the $(g/4)|\psi|^4$ interaction term in (106). An increase in $m_A$ corresponds to a decrease of the penetration depth in the superconductor, i.e. to materials moving towards the type-I regime. At some place, the $|\psi|^4$-term will vanish and the disorder field theory requires a $|\psi|^4$-term to stabilize the fluctuations of the vortex lines. In such materials, the superconductive phase transition turns from second to first order. The existence and position of such a tricriticsl point was first established in [10].

### 2.11. Vortex Lattices

The model action (14) represents the gradient energy in superfluid $^4$He correctly only in the long-wavelength limit. The neutron scattering data yield the energy spectrum $\omega = S(\mathbf{k})$ shown in Fig. 3. To account for this, the action should be taken as follows:

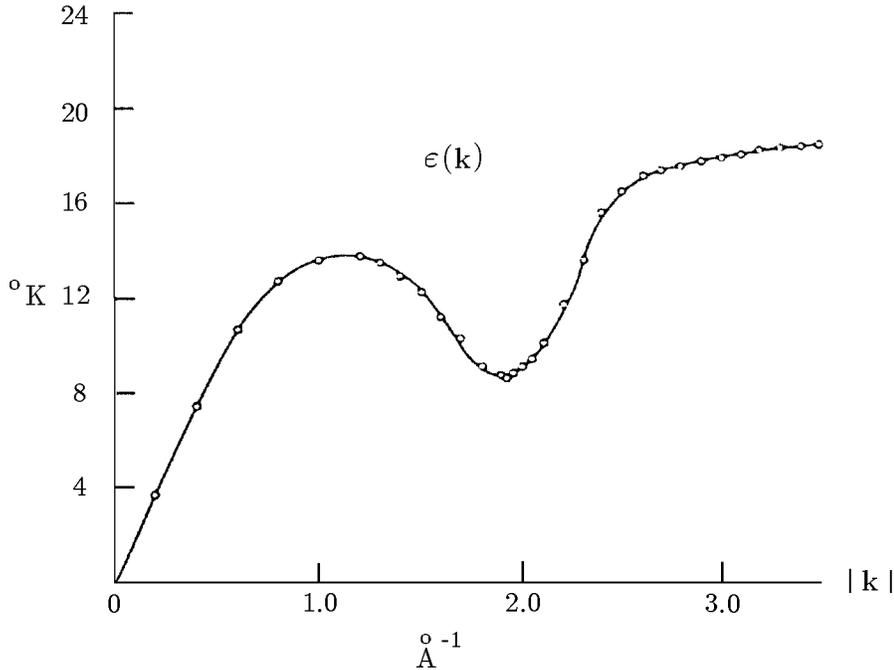

**Figure 3.** The structure factor of superfluid $^4$He measured by neutron scattering showing the excitation energy of the NG-bosons.

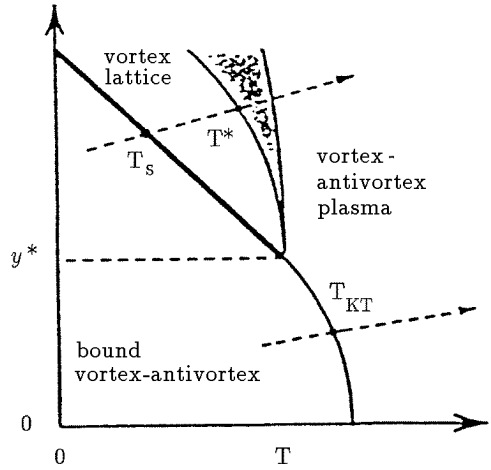

**Figure 4.** The phase diagram of a two-dimensional layer of superfluid $^4$He. At higher fugacity, the vortices first condense to a lattice, which undergoes the Kosterlitz-Thouless vortex unbinding transition only after a melting transition.

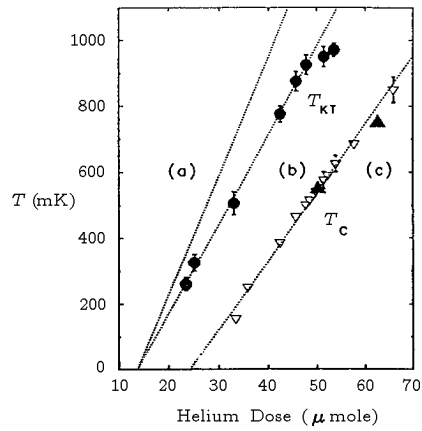

**Figure 5.** The experimental phase diagram of a two-dimensional layer of superfluid $^4$He diluted with $^3$He which decreases the fugacity and separates the vortex melting transition from the Kosterlitz-Thouless transition.

$$\mathcal{A}_{\mathrm{NG}} = \frac{1}{2T} \int d^3x (\partial\theta - \boldsymbol{\theta}^{\mathrm{v}}) \frac{S(-i\partial)}{\mathbf{k}^2}(\partial\theta - \boldsymbol{\theta}^{\mathrm{v}}). \tag{140}$$

The *roton peak* near 20Å$^{-1}$ gives rise to a repulsion between opposite vortex line elements at the corresponding distance. If a layer of superfluid $^4$He is diluted with $^3$He, the core energy of the vortices decreases, the fugacity $y$ and the average vortex number increases. For a sufficiently high average spacing, a vortex lattice forms. In this regime, the superfluid has three transition when going from zero temperature zero to the normal phase, first condensation process of a vortex lattice, then a melting transition of this lattice into a fluid of bound vortex-antivortex pairs, and third a pair-unbinding transitions of the Kosterlitz-Thouless type [11, 12]. The latter two transitions have apparently been seen experimentally [13].

### 3. Crystals

In crystals, the elastic energy is usually expressed in terms of a material *displacement field* $u_i(\mathbf{x})$ as

$$E = \int d^3x \, u_{ij}^2(\mathbf{x}) \tag{141}$$

where

$$u_{ij}(\mathbf{x}) = \frac{1}{2}(\partial_i u_j(\mathbf{x}) + \partial_j u_j(\mathbf{x})) \tag{142}$$

is the *strain tensor*. For simplicity, we have set the shear elastic constant $\mu$ equal to unity and neglected the second elastic term $\lambda(\partial_i u_i)^2$. The elastic energy goes to zero for infinity wave length since in this limit $u_i(\mathbf{x})$ reduces to a pure translation and the energy of the system is translationally invariant. The crystallization process causes a spontaneous breakdown of the translational symmetry of the system elastic distortions are $NG$-modes of this symmetry breakdown.

A crystallize material always contains defects. In their presence, the elastic energy is

$$E = \int d^3x \, (u_{ij} - u_{ij}^{\mathrm{p}})^2 \tag{143}$$

where $u_{ij}^{\mathrm{p}}$ is the so-called *plastic strain tensor* describing the defects. Topologically, the crystal defects arise as a consequence of the multivaluedness of the displacement field on a lattice. Due to the identity of the lattice constituents, the displacement is defined only modulo lattice vectors. For mathematical simplicity, let us assume the lattice to be simple cubic of spacing $2\pi$. Then the energy of $u_i(\mathbf{x})$ and $u_i(\mathbf{x}) + 2\pi N_i(\mathbf{x})$ must be indistinguishable for integer valued fields $N_i(\mathbf{x})$ which correspond to a permutation of the lattice sites. The order field of a crystal has therefore the three components

$$\psi_i(\mathbf{x}) = e^{i2\pi u_i(\mathbf{x})} \psi_{0i}. \tag{144}$$

This is very similar to the order field of a superfluid. There are now three types of surfaces where the *distortion tensor* $\beta_{il} \equiv \partial_i u_l$ jumps by $2\pi$, one for every lattice direction $b_l = 1, 2, 3$. They are describes by the *plastic distortion*

$$\beta_{il}^{\mathrm{P}}(\mathbf{x}) = \delta_i(\mathbf{x}; S) b_l. \tag{145}$$

The surfaces $S$ are irrelevant and correspond to a layer of atoms missing from a crystal. Only the boundary line is physical. They are called *dislocation lines* of Burger vector dislocation lines are extracted from $\beta_{ij}^{\mathrm{P}}$ by forming the dislocation density

$$\alpha_{il}(\mathbf{x}) = \epsilon_{imn}\partial_m \beta_{nl}^{\mathrm{P}}(\mathbf{x}) = \delta_i(L)b_l. \tag{146}$$

Being a curl, it satisfies the conservation law

$$\partial_i \alpha_{il} = 0 \tag{147}$$

trivially as a Bianchi identity. The conservation law states that dislocation lines are closed.

A closed Volterra surface is associated with a translation

$$u_i(\mathbf{x}) \to u_i(\mathbf{x}) + 2\pi b_i \delta(\mathbf{x}; V) \tag{148}$$

where $V$ is the volume enclosed by the closed Volterra surface $S$. Under such an operation, the order parameters $e^{iu_i(\mathbf{x})}$ are invariant.

In a real crystal, dislocation lines are able to stack up on top of each other forming line-like defects called *disclination lines*. Their geometric properties are most easily understood by noting that instead of the trivial translation (148) across a closed Volterra surface we can form a combined translation rotation plus

$$u_i(\mathbf{x}) \to [b_i + R_{ij}u_j(\mathbf{x})]\,\delta(\mathbf{x}; V) \tag{149}$$

where $R_{ij}$ is a discrete rotation under which the crystal is invariant (in a simple cubic lattice, $R_{ij}$ consists of products of rotations by $\pi/2$). The nonabelian nature of the trivial gauge transformations (149) makes a full discussion of the plastic deformations quite complicated. A useful approximation consists in assuming $R_{ij}$ to be a small rotation and to reexpress (149) in the infinitesimal form

$$u_l(\mathbf{x}) \to u_l(\mathbf{x}) + (b_l + \epsilon_{lqn}\Omega_q x_r)\delta(\mathbf{x}; V). \tag{150}$$

Assuming the presence of a closed Volterra surface, we extract the jumps in the distortion field

$$\beta_{kl} \equiv i\partial_k u_l(\mathbf{x}) = \delta_k(\mathbf{x}; S)(b_l + \epsilon_{lqr}\Omega_q x_r) - \delta(\mathbf{x}; V)\epsilon_{lqk}\Omega_q. \tag{151}$$

The first term is defined also for an open surface in which case it becomes the plastic distortion of a general defect line

$$\beta_{kl}^{\mathrm{P}} = \delta_k(\mathbf{x}; S)(b_l + \epsilon_{lpr}\Omega_q x_r). \tag{152}$$

By forming the curl of the distortion field (151), the term $\delta(\mathbf{x}; V)$ drops out and we find

$$\alpha_{il} = \epsilon_{ijk}\partial_j \partial_k u_l(\mathbf{x}) = \delta_i(\mathbf{x}; L)(b_l + \epsilon_{lqr}\Omega_q x_r). \tag{153}$$

This is called the *dislocation density* of the defect line. The rotation field

$$\omega_j = \frac{1}{2}\epsilon_{jkl}\partial_u u_l \tag{154}$$

has the following derivative

$$\partial_n \omega_j = \frac{1}{2}\epsilon_{jkl}\partial_k \partial_n u_l = \frac{1}{2}\epsilon_{jkl}\partial_k \beta_{kl}^{\mathrm{P}} + \delta_n(\mathbf{x}; S)\Omega_j. \tag{155}$$

$$\phi^{\mathrm{P}}_{nj} = \delta_n(s)\Omega_j. \tag{156}$$

The entire tensor (155) is called the *plastic bend-twist*

$$\kappa^{\mathrm{P}}_{nj} \equiv \partial_n \omega_j = \frac{1}{2}\epsilon_{ikl}\partial_n \beta^{\mathrm{P}}_{kl} + \phi^{\mathrm{P}}_{nj}. \tag{157}$$

The curl of the plastic rotation is the *disclination density*

$$\theta_{il} = \epsilon_{ijk}\partial_j \phi^{\mathrm{P}}_{kl} = \delta_i(\mathbf{x}; L)\Omega_l. \tag{158}$$

Being a curl, this satisfies the conservation law

$$\partial_i \theta_{il} = 0 \tag{159}$$

implying that disclination lines are closed. This is again a Bianchi identity of the defect system. The conservation law for dislocations is modified by the presence of disclination as follows:

$$\partial_j \alpha_{ik} = -\epsilon_{imn}\theta_{mn}. \tag{160}$$

Dislocation lines can now end in disclination lines.

An important geometric quantity characterizing dislocation and disclination lines is the *incompatibility* or *defect density*

$$\eta_{ij}(\mathbf{x}) = \epsilon_{ikl}\epsilon_{jmn}\partial_k \partial_m u_{ln}(\mathbf{x}). \tag{161}$$

It can be decomposed into disclination and dislocation density as follows:

$$\eta_{ij}(\mathbf{x}) = \theta_{ij}(\mathbf{x}) + \frac{1}{2}\partial_m \left[\epsilon_{min}\alpha_{jn}(\mathbf{x}) + (ij) - \epsilon_{ijn}\alpha_{mn}(\mathbf{x})\right]. \tag{162}$$

This tensor is symmetric and conserved

$$\partial_i \eta_{ij}(\mathbf{x}) = 0, \tag{163}$$

again a Bianchi identity of the defect system.

The tensors $\alpha_{ij}, \theta_{ij}$ and $\eta_{ij}$ are linearized versions of important tensors in a *Riemann-Cartan space*, a noneuclidean space with curvature and torsion. Such a space can be generated from a flat space by a plastic distortion, which is mathematically represented by a *nonholonomic* mapping

$$x_i \to x_i + u_i(\mathbf{x}). \tag{164}$$

Such a mapping is nonintegrable. The displacement fields and their first derivatives fail to satisfy the Schwarz integrability criterion:

$$\begin{aligned}(\partial_i \partial_j - \partial_j \partial_i)u(\mathbf{x}) &\neq 0 \\ (\partial_i \partial_j - \partial_j \partial_i)\partial_k u_l(\mathbf{x}) &\neq 0.\end{aligned} \tag{165}$$

The affine connection of the geometry in the plastically distorted space is

$$\Gamma_{ijl} = \partial_i \partial_j u_l. \tag{166}$$

torsion tensor being
$$S_{klj} \equiv (\Gamma_{ijk} - \Gamma_{jik})/2. \tag{167}$$

The dislocation density $\alpha_{ij}$ is equal to
$$\alpha_{ij} = \epsilon_{ikl} S_{klj}, \tag{168}$$

The noncommutativity of the derivatives in front of $\partial_k u_l(\mathbf{x})$ imply a nonzero curvature. The disclination density $\theta_{ij}$ is the Einstein tensor
$$\theta_{ij} = G_{ij} \equiv_{ji} -\frac{1}{2} g_{ji} R \tag{169}$$

of this geometry. The tensor $\eta_{ij}$, finally, is the Belinfante symmetric energy momentum tensor which coincides with the Einstein tensor obtained from the Christoffel symbols of the metric
$$g_{ij} = \delta_{ij} + \partial_i u_j + \partial_j u_i \tag{170}$$
rather than from the affine connection $\Gamma_{ijk}$. To verify this statement we just note that the spin current density of the gravitational field is (setting the gravitational constant equal to unity)
$$\Sigma_{ij,k} = -2 \left( S_{ijk} + \delta_{ik} S_j - \delta_{jk} S_i \right). \tag{171}$$
For more details on the geometric aspects see Vol. II of [1], Part IV, where the full one-to-one correspondence between defect systems an Riemann-Cartan geometry is developed as well as a gravitational theory based on this analogy.

It is possible to write down an elastic energy which disentangles dislocations and disclinations by including higher gradients of the displacements field. This energy reads
$$E = \int d^3x \left[ \left( u_{ij} - u_{ij}^{\mathrm{P}} \right)^2 + \ell^2 \left( \partial_i \omega_j - \kappa_{ij}^{\mathrm{P}} \right)^2 \right]. \tag{172}$$

The parameter $\ell$ is the length scale over which the crystal is rotationally stiff.

The partition function contains integrals over $u_i$ and sums over the jumping surfaces of dislocations and disclinations. By integrating out the $u_i$-fields, one obtains a Biot-Savart type of interaction energies between the defect lines in which dislocation line elements interacts with each other via a Coulomb potential, and disclination line elements via an $r$-potential.

It is again possible to eliminate the jumping surfaces from the partition function by introducing conjugate variables and associated stress gauge fields. For this we rewrite the elastic action of defect lines as
$$E = \int d^3x \left[ \frac{1}{4T} (\sigma_{ij} + \sigma_{ji})^2 + \frac{1}{8\ell^2} \tau_{ij}^2 + i\sigma_{ij} \left( \partial_i u_j - \epsilon_{ijk} \omega_k - \beta_i^{\mathrm{P}} \right) + i\tau_{ij} \left( \partial_i \omega_j - \phi_{ij}^{\mathrm{P}} \right) \right], \tag{173}$$
and form the partition function by integrating over $\sigma_{ij}, \tau_{ij}, u_i, \omega_j$ and summing over all jumping surfaces $S$ in the plastic fields. A functional integral over the antisymmetric part of $\sigma_{ij}$ has been introduce the obtain an independent integral over $\omega_i$ (if we were to integrate out the antisymmetric part of $\sigma_{ij}$, we would enforce the relation $\omega_i = \frac{1}{2} \epsilon_{ijk} (\partial_j u_k + \beta_{ij}^{\mathrm{P}})$. By integrating out $\omega_j$ and $u_i$, we find the conservation laws
$$\partial_i \sigma_{ij} = -\epsilon_{jkl} \tau_{kl}, \quad \partial_i \tau_{ij} = 0. \tag{174}$$

and (159).

They can be enforced as Bianchi identities by introducing the stress gauge fields $A_{ij}$ and $h_{ij}$ and writing

$$\begin{aligned} \sigma_{ij} &= \epsilon_{ikl}\partial_k A_{lj} \\ \tau_{ij} &= \epsilon_{ikl}\partial_k h_{lj} - \delta_{ij}A_{ll} - A_{ji}. \end{aligned} \quad (175)$$

This allows to reexpress the energy as

$$E = \int d^3x \left[ \frac{1}{4}(\sigma_{ij} + \sigma_{ji})^2 + \frac{1}{8l^2}\tau_{ij}^2 + A_{ij}\alpha_{ij} + h_{ij}\theta_{ij} \right]. \quad (176)$$

The stress gauge fields couple locally to the defect densities which are singular on the boundary lines of the Volterra surfaces. In the limit of a vanishing length scale $\ell$, $\tau_{ij}$ is forced to be identically zero and (175) allows us to express $A_{ij}$ in terms of $h_{ij}$. Then the energy becomes

$$E = \int d^3x \left[ \frac{1}{4}(\sigma_{ij} + \sigma_{ji})^2 + h_{ij}\eta_{ij} \right] \quad (177)$$

where the defect density $\eta_{ij}$ contains dislocation and disclination lines.

Depending on the length parameter $\ell$ of rotational stiffness, the defect system was predicted to have either a single first-order transition (for small $\ell$), of two successive continuous melting transitions. In the first transitions, dislocation lines proliferate and destroy the translational order, in the second transition, disclination lines proliferate and destroy the rotational order [4].

In two dimensions, the existence of such a scenario (where the continuous transitions for small $\ell$ would be of the Kosterlitz-Thouless type) was conjectured a long time ago [14, 15], but the defect models constructed to illustrate this displayed only a single first-order transition [16]. The computer simulations of the model containing the length parameter $\ell$ [17], on the other hand, confirmed the theoretical expectation citekl2tr.

### 4. Abelian Quark Confinement

In Section II we have observed the appearance of confining forces caused by the breakdown of vortex gauge invariance: An irrelevant geometric line becomes energetic and creates a linearly rising potential between the endpoints. This mechanism can be used in four spacetime dimensions to construct an abelian theory of quark confinement. In a superconductor, whose ground state contains a condensate of electric charges, magnetic monopoles are confined. To obtain confining force between electric charges, we have to construct a dual theory containing poliferating monopole world lines.

The euclidean field action of such world lines is given by

$$\mathcal{A} = \frac{1}{4}\int d^4x \left\{ [F_{\mu\nu}(\mathbf{x}) - F^{\rm m}_{\mu\nu}(\mathbf{x})]^2 + ij_\mu(\mathbf{x})A_\mu(\mathbf{x}) \right\} \quad (178)$$

where $F_{\mu\nu} = \partial_\mu A_\nu - \partial_\nu A_\mu$ is the usual electromagnetic field tensor,

$$j_\mu(\mathbf{x}) = e\delta_i(\mathbf{x}; L) \quad (179)$$

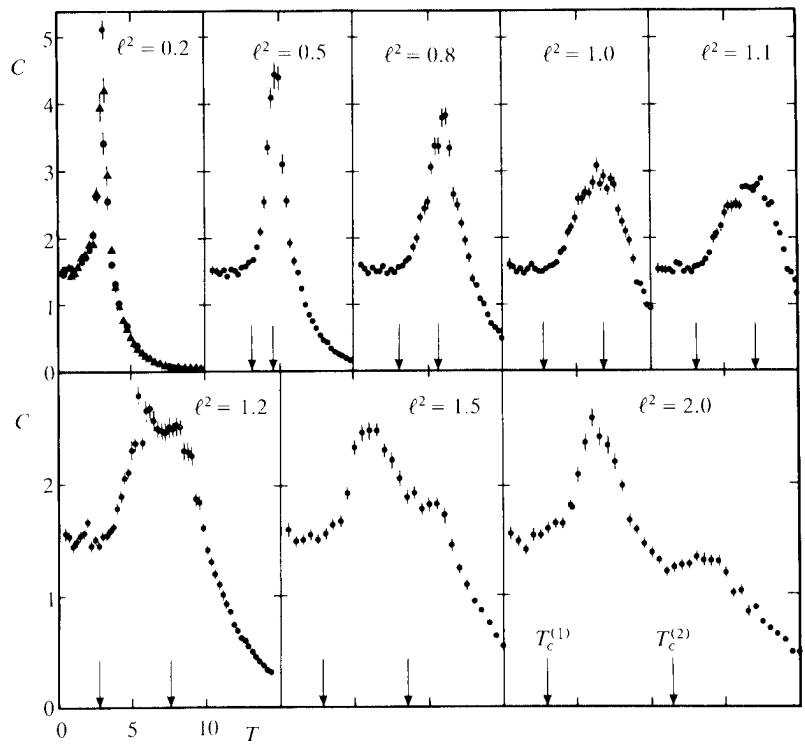

**Figure 6.** Separation of first-order melting transition into two successive Kosterlitz-Thouless transitions when increasing the length scale $\ell$ of rotational stiffness of the defect model.

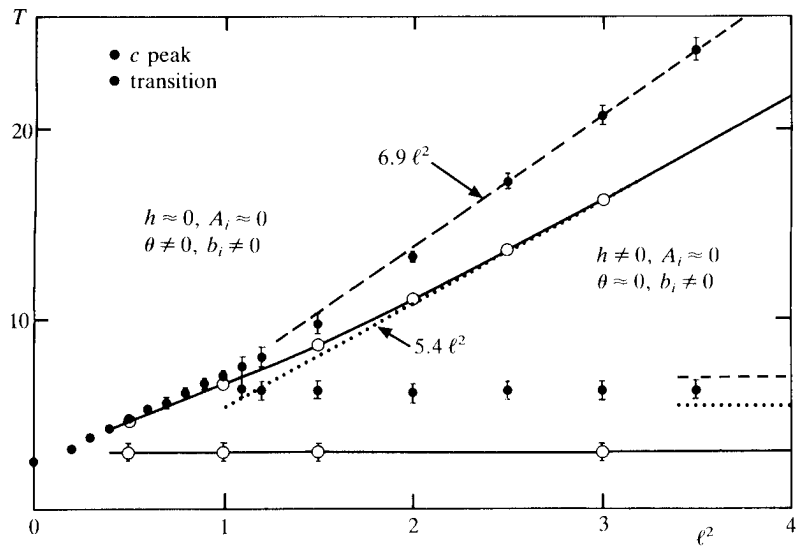

**Figure 7.** The phase diagram in the $T$-$\ell$-plane.

$$F_{\mu\nu}^{\mathrm{m}} = g\frac{1}{2}\epsilon_{\mu\nu\lambda\kappa}\delta_{\lambda\kappa}(\mathbf{x};\tilde{S}) \tag{180}$$

is the *gauge field of monopoles* . The $\delta$-function on the surface $\tilde{S}$

$$\delta_{\lambda\kappa}(\mathbf{x};\tilde{S}) = \int d\sigma d\tau \left[\frac{\partial \bar{x}_\mu}{\partial \sigma}\frac{\partial \bar{x}_\nu}{\partial \tau} - (\mu\nu)\right] \delta^{(4)}(\mathbf{x} - \bar{\mathbf{x}}(\sigma,\tau)) \tag{181}$$

has the property that

$$\partial_\kappa \delta_{\lambda\kappa}(\mathbf{x};\tilde{S}) = \delta_\lambda(\mathbf{x};\tilde{L}) \tag{182}$$

where $\tilde{L}$ is the boundary line of $\tilde{S}$. Hence

$$\frac{1}{2}\epsilon_{\mu\nu\lambda\kappa}\partial_\nu F_{\lambda\kappa}^{M} = \tilde{j}_\mu(\mathbf{x}) \tag{183}$$

where

$$\tilde{j}_\mu(\mathbf{x}) = g\delta_\mu(\mathbf{x};\tilde{L}) \tag{184}$$

is the magnetic current of the monopole world line along $\tilde{L}$. The field energy invariant under monopole gauge transformations $\tilde{S} \to \tilde{S}'$ under which

$$\delta_{\lambda\kappa}(\mathbf{x};\tilde{S}) \to \delta_{\lambda\kappa}(\mathbf{x};\tilde{S}) + \epsilon_{\lambda\kappa\sigma\tau}\partial_\sigma \delta_\tau(\mathbf{x};V) \tag{185}$$

where

$$\delta_\tau(\mathbf{x};\tilde{V}) = \epsilon_{\mu\nu\kappa\gamma}\int d\sigma d\tau\lambda \frac{\partial \bar{x}_\nu}{\partial \sigma}\frac{\partial \bar{x}_\mu}{\partial \tau}\frac{\partial \bar{x}_\lambda}{\partial_\lambda}\delta^{(n)}(\mathbf{x} - \bar{\mathbf{x}}(\sigma,\tau,\lambda)) \tag{186}$$

is the $\delta$-function on the volume over which $\tilde{S}$ has swept. The vortex gauge field transforms as follows

$$F_{\mu\nu}^{\mathrm{m}} \to F_{\mu\nu}^{\mathrm{m}} + \partial_\mu \Lambda_\nu^{\mathrm{m}} - \partial_\nu \Lambda_\mu^{\mathrm{m}} \tag{187}$$

where

$$\Lambda_\mu^{M} = 4\pi q\delta(\mathbf{x};\tilde{V}). \tag{188}$$

The gauge functions $\Lambda_\mu^{\mathrm{m}}$ can be absorbed into the vector potential

$$A_\mu \to A_\mu + \Lambda_\mu^{M}. \tag{189}$$

The additional interaction term $\int d^4x j_\mu(\mathbf{x})\Lambda_\mu^{\mathrm{m}}(\mathbf{x})$ changes by $eg$ times the integral $\int d^4x \delta_\mu(\mathbf{x};L)\delta_\mu(\mathbf{x};\tilde{V})$. The integral is an integer number $n$ counting how many times the line $L$ pierces the volume $\tilde{V}$. Thus, the functional integral of the system

$$Z = \int \mathcal{D}A \sum_{\tilde{S}} e^{-\mathcal{A}} \tag{190}$$

picks up a phase factor $e^{iegn}$. This is an irrelevant factor if

$$eg = 2\pi n, \tag{191}$$

which is Dirac's charge quantization condition.[4] By introducing an auxiliary field $\tilde{f}_{\mu\nu}$, the theory can be transformed to the dual form

$$\mathcal{A} = \int d^4x \left[\frac{1}{2}\left(\tilde{F}_{\mu\nu} - \bar{F}_{\mu\nu}^{\mathrm{e}}\right)^2 + i\int \bar{j}_\mu \bar{A}_\mu\right]. \tag{192}$$

---

[4]Dirac uses a field energy with a prefactor $1/4\pi$ in which case $e$ and $g$ have to be replaced by $\sqrt{4\pi}e$ and $\sqrt{4\pi}g$ and the condition reads $2eg = n$.

is locally coupled to the magnetic current $\tilde{j}_\mu(\mathbf{x})$. The world line $L$ of the electric charges are now represented by a gauge field

$$\tilde{F}^{\rm e}_{\mu\nu} = e\frac{1}{2}\epsilon_{\mu\nu\lambda\kappa}\delta_{\lambda\kappa}(\mathbf{x};S) \tag{193}$$

with some irrelevant surface $S$ whose boundary is $L$. This field energy is invariant under the dual charge gauge transformations

$$\begin{aligned}\tilde{F}^{\rm e}_{\mu\nu} &\to \tilde{F}^{\rm e}_{\mu\nu} + \partial_\mu\tilde{\Lambda}^{\rm e}_\nu - \partial_\nu\tilde{\Lambda}^{\rm e}_\mu,\\ \tilde{A}_\mu &\to \tilde{A}_\mu + \tilde{\Lambda}^{\rm e}_\mu.\end{aligned} \tag{194}$$

At higher temperatures, the monopole world lines proliferate on account of their entropy. In this phase, they can be described by a disorder field theory and the total action reads

$$\mathcal{A} = \int d^4x\left\{\frac{1}{4}\left(\tilde{F}_{\mu\nu} - \tilde{F}^{\rm e}_{\mu\nu}\right)^2 + \frac{\tilde{m}_A^2}{2g^2}\left(\partial_\mu\theta - g\tilde{A}_\mu\right)^2\right\}. \tag{195}$$

By integrating out the angular variables $\theta$ of disorder, we are left with a transverse mass term of the $\tilde{A}_\mu$ field

$$\int d^4x \frac{\tilde{m}_A^2}{2}A_\mu^{T^2}. \tag{196}$$

As a consequence of this, the previously irrelevant surfaces $S$ of the charge gauge fields become massive and give rise to a continuing force between the boundary lines. Going back to the electromagnetic description, the action can be written as

$$\mathcal{A} = \int d^4x\left\{\frac{1}{4}(F_{\mu\nu} - f_{\mu\nu})^2 + ij_\mu A_\mu - i\tilde{A}_\mu\left(\frac{1}{2}\epsilon_{\mu\nu\lambda\kappa}\partial_\nu f_{\lambda\kappa} - \tilde{j}_\mu\right) + \frac{1}{2\tilde{m}_A^2}\tilde{j}_\mu^2\right\}. \tag{197}$$

Here $\tilde{A}_\mu$ plays the role of a Lagrangian multiplyer which enforces $f_{\mu\nu}$ to be a gauge field for the monopole world lines $\tilde{L}$.

In the high-temperature phase, the monopoles proliferate and $\tilde{j}_\mu$ can be treated as an ordinary fluctuating transverse field giving rise to the mass term $\frac{\tilde{m}_A^2}{2}\tilde{A}_\mu$. After this, $\tilde{A}_\mu$ can be integrated out and produces a term $1/2\tilde{m}_A^2\left(\frac{1}{2}\epsilon_{\mu\nu\lambda\kappa}\partial_\nu f_{s\kappa}\right)^2$. The $f_{\mu\nu}$-field is responsible for the generation of the mass term of the surfaces $S$ in the gauge field description of the electric charges, as can be verified by going again through a duality transformation.

Near the critical temperature, the fluctuating $\tilde{j}$-currents can be represented by a disorder field theory of the $|\psi|^4$-type and the total action reads

$$\begin{aligned}\mathcal{A} &= \int d^4x\left\{\frac{1}{4}(F_{\mu\nu} - f_{\mu\nu})^2 + ij_\mu A_\mu - i\tilde{A}\left(\frac{1}{2}\epsilon_{\mu\nu\lambda\kappa}\partial_\nu f_{\lambda\kappa} - \tilde{j}_\mu\right)\right.\\ &\quad\left. + |\left(\partial_\mu - i\tilde{A}_\mu\right)\psi|^2 + m^2|\psi|^2 + \frac{g}{2}|\psi|^4\right\}.\end{aligned} \tag{198}$$

## ACKNOWLEDGMENTS

I thank Dr. A. Schakel and M. Kiometzis for useful discussions.